\documentclass[aps,notitlepage,superscriptaddress,prl,twocolumn]{revtex4-2}
\usepackage{ucs}
\usepackage[utf8x]{inputenc}
\usepackage{amsmath}
\usepackage{amsfonts}
\usepackage{amssymb}
\usepackage{amsthm}
\usepackage{fontenc}
\usepackage[dvips,pdftex]{graphicx}
\usepackage{setspace}
\usepackage{color}
\usepackage[dvipsnames]{xcolor}
\usepackage{graphicx}

\usepackage{hyperref}

\usepackage{multirow}
\usepackage{array}
\usepackage{subcaption}
\usepackage{tikz}
\usepackage{multimedia}
\usepackage{float}
\usepackage{rotating}
\usepackage{epstopdf}
\usepackage{soul}

\makeatletter
\@addtoreset{chapter}{part}
\makeatother

\newcommand{\dt}{\,\text{d}}

\newcommand{\pd}{\partial}
\newcommand{\Tr}{\text{Tr}}

\newcommand{\lpow}[2]{\,{}^{#1}\!#2}

\newcommand{\ptwo}{\lpow{2}{P}}

\newcommand{\Qstar}{{Q^{*}}}

\DeclareGraphicsExtensions{.pdf,.eps,.jpg,.png}

\hypersetup{linktocpage}

\begin{document}

	\title{Predicting the mechanical properties of spring networks}
	\date{\today}
	
	\author{Doron \surname{Grossman}}
	\email[]{doron.grossman@ladhyx.polytechnique.fr}
	\affiliation{College de France, 11 Pl. Marcelin Berthelot, 75231 Paris}
	\author{Jean-Francois \surname{Joanny}}
	\affiliation{College de France, 11 Pl. Marcelin Berthelot, 75231 Paris}
	\date{\today}

%
%
%


\begin{abstract}
 {The mechanical properties of tissues play  an essential role for all tissue properties such as  cell division, and differentiation or morphogenesis. Here,  we study theoretically  the rheology of 2-dimensional epithelial tissues described by a discrete vertex-like model, using an analytical coarse-grained continuum formulation}. We show that epithelial tissues are {most often} shear-thinning under constant shear rate, and {in certain circumstances cross over from} shear thickening at low shear rates 
 to shear thinning at high shear rates. We  give  {an} analytical expression {of the}  tissue response in an  oscillating strain experiment {in the} linear regime, and {calculate it numerically in the non-linear regime}. When the tissue is supported by an oscillating substrate, it reorients depending on frequency and substrate's Poisson's ratio. Reorientation could be gradual or abrupt, depending on tissue and substrate parameters, and the configuration phase space exhibits a tricritical point.
\end{abstract}


\maketitle

%
{A} biological living tissue subject to external  stresses or strains, behaves as an active viscoelastic or visco-plastic material, within which various cellular processes allow for the creation and dissipation of internal stresses. The exact nature of the relevant dissipative processes depends on the type of tissue as well as on  the interaction with its environment. Plant tissues behave as viscoelastic solids, in which turgor pressure and osmotic balance within the tissue play a significant role in setting both the shape and the mechanical response to deformations \cite{mirabet2011role,goriely2017mathematics}. 
Animal tissues, on the other hand, behave as viscoelastic or viscoplastic  fluids, for which stress relaxation is driven by cell
proliferation (division and apoptosis) and  relative cellular rearrangements \cite{blanch2017effective}.  

Vertex models are  discrete models often used to describe all types of cellular media, in which cells are described as confluent polygons in two dimensions or polyhedra in 3 dimensions. \cite{thompson1942growth,ranft2010fluidization,hannezo2014theory,farhadifar2007influence,staple2010mechanics,ishihara2017cells,tong2023linear}. They have been used very successfully to describe the behavior of squamous epithelia  \cite{Etournay2015,guirao2015unified}. In this paper we study the rheological properties of tissues described by the generalized  2-dimensional vertex model that we introduced in Ref.~\cite{grossman2021instabilities}.  In the standard vertex model, each cell has a preferred area $A^0_i$ and a preferred perimeter $P^0_i$. In the generalized model we replace the perimeter  $P = \sum_e \ell_e$ (sum of the lengths of the edges of a cell) by  the generalized perimeter, $\ptwo = \sum_e \ell^2_e$ (sum of the squared lengths of the edges of each cell), in order to allow for simpler analytical calculations, in a rigorous and justified manner. Each cell has then a preferred generalized perimeter $\ptwo^0_i$. The energy of the generalized vertex model has the same form as that of the original formulation
\begin{align}\label{eq: vertex model 1}
	E & =  \sum_i  \hat{k}_A \left(A_i - \hat{A}^0_i\right)^2 + k_p \left(\ptwo_i -\ptwo^0_i\right)^2.
\end{align}
where $\hat{k}_A$ and $ k_p$ are positive area and perimeter moduli and $A_i$ and $\ptwo_i$ the actual area and perimeter of a given cell indexed by $i$. We checked explicitly in 
Ref.~\cite{grossman2021instabilities} that  the generalized vertex model leads quantitatively to
the same behavior 
as the standard vertex model at linear order in the differences of area and perimeter with the reference 
area and perimeter and  to a similar qualitative behavior beyond linear order, using examples where the tissue behavior is known.

In {Ref.~}\cite{grossman2021instabilities} we also developed a coarse-grained continuum formulation of our generalized vertex 
model in a hydrodynamic limit where the locally averaged cell area and perimeter vary over distances that are much 
larger than the cell size. In the mean field approximation where we ignore terms containing gradients of the average area 
and perimeter,  we obtain the continuum energy as a function of two tensors defined locally. 
The metric tensor $\bf g$ (with components $g_{\mu\nu}$, and determinant $g$) prescribes the distance within the tissue using a reference 
state that evolves with 
time because of the topological transitions in the tissue. The network tensor, $\bf Q$ (with components $Q^{\mu\nu}$, and determinant $Q$) is defined in the 
reference state and prescribes the local cellular shape.  Formally, $Q^{\mu\nu}$ is given by -  
\begin{align}
	Q^{\mu\nu}= \langle \sum_e \Delta x_e^{\mu} \Delta x_e^{\mu}\rangle = \frac{1}{N_{cells}}\sum_i \sum_{e \in i }  \Delta x_e^{\mu} \Delta x_e^{\mu}
\end{align} where the triangular parentheses {indicate} a local average {defined on the right hand side} {where}  $N_{cells}$ is the number of cells in the region under averaging,  $i$ is a cell index $e\in i$ is an edge index {of} the $i^{th}$ cell, $\Delta x_e^\mu$ is the $\mu$ coordinate  {of the vector joining the two vertices of} the edge $e$. The energy {of the tissue} is then
\begin{align}\label{eq: avg energy}
	E = \int \rho(x) \Big\{ & k_A \left( \sqrt{g Q} - A^0\left( x \right) \right)^2   \\\nonumber & +k_p \left(g_{\mu\nu}Q^{\mu\nu} - 
	\ptwo^0\left( x \right) \right)^2  \Big\}\sqrt{g} \dt x .
\end{align}
 After rescaling numerical prefactors, 
the area of a cell is approximated by $A= \sqrt{gQ}$ and the perimeter of a cell is the trace $\ptwo=g_{\mu\nu}Q^{\mu\nu}$. The 
rescaling changes the area modulus to a value $k_A$. Changes in the metric describe how the tissue as a whole deforms and occupies space (elastic deformation), while changes in the network tensor indicate how the tissue's internal structure changes, i.e how the cells rearrange (plastic deformation). 

One of the important predictions of the standard vertex model 
\cite{farhadifar2007influence,moshe2018geometric} is the existence of a solid-solid transition between a hard isotropic solid tissue and a soft oriented solid tissue at a critical shape parameter $p=\frac{P^0}{\sqrt A^0} = 3.722$. 
Due to our reformulation, this transition occurs at a shape parameter  $p=\ptwo^0/A_0=2$, if fluctuations of area and perimeter are ignored.  
When the shape parameter is larger than $2$, cells have an elongated shape and the tissue is 
completely relaxed without any residual internal stresses. When the shape parameter is smaller than $2$, cells are 
isotropic but the tissue has residual internal stresses. The existence of residual stresses plays an 
essential role for the mechanical properties of the tissue. In the following, we fix the length scale and the energy scale by imposing $A_0=1$ and $k_A=1$.

In this paper, we study  the linear and the non-linear rheology of a 2 dimensional  tissue described by the generalized vetex model in two different 
geometries. We first study a free standing tissue with an imposed shear rate both when the shear rate is 
constant and when it oscillates at a finite pulsation $\omega$. In a second part,  we study a tissue 
supported by an elastic substrate, with a Poisson 
ratio $\nu$,  stretched and relaxed in one direction in an oscillatory manner \cite{geremie2022evolution, Lucci2021}.  

All of the results presented here are new in the context of vertex models, and could not be calculated analytically without the continuum model we developed in \cite{grossman2021instabilities}. The continuum model used here is valid as long as cell size is smaller than the scale over which the metric changes.

\section{Sheared tissue}\label{section: shear}

\begin{figure}
	\centering
	\includegraphics[width=\columnwidth]{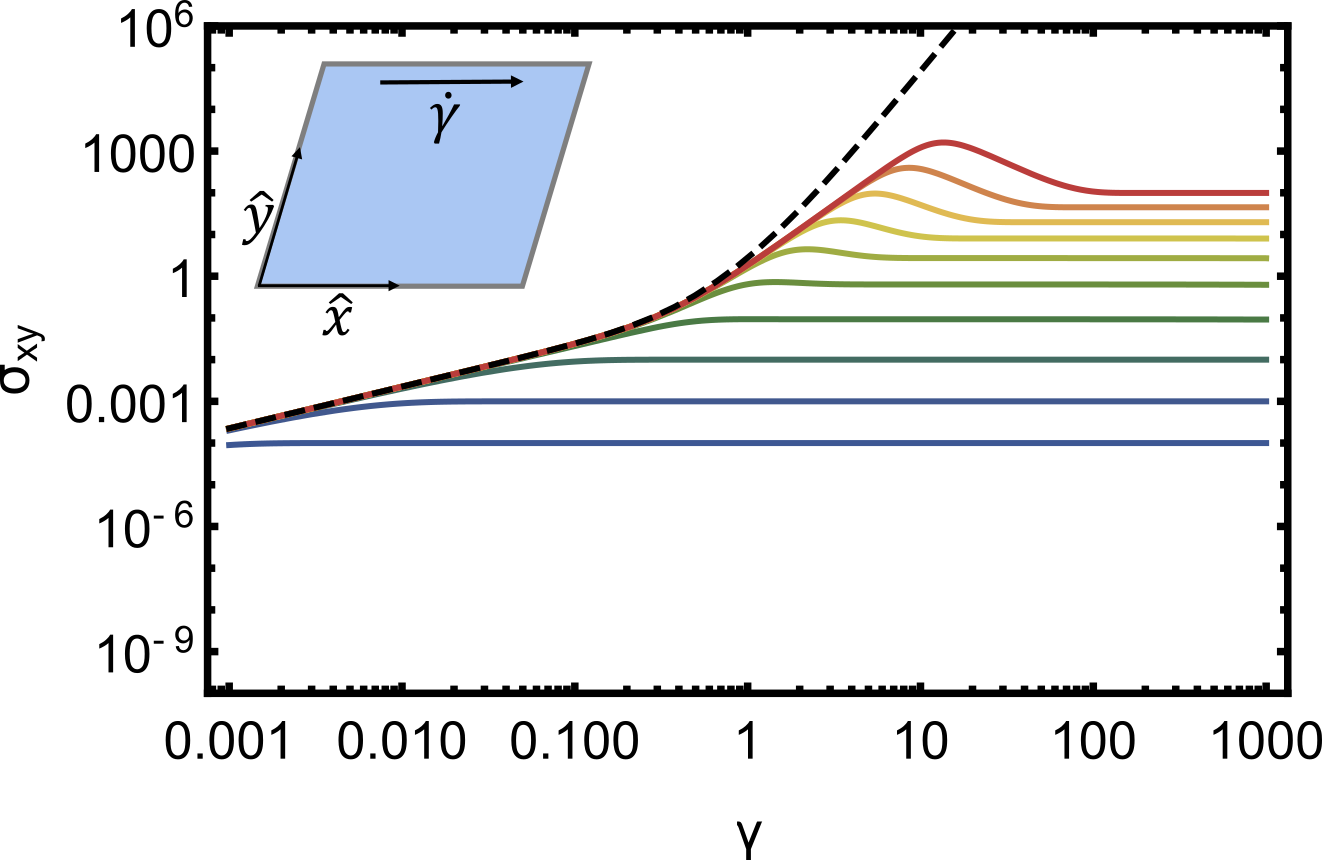}
	\caption{Shear stress as a function of strain, at constant shear rate, when $p=1.5<2$ and $H= -1/2 $.  The  colored lines correspond to different shear rates:  blue (bottom plot) - $\dot\gamma = 10^{-4}$, red (top plot) - $\dot\gamma = 10^5$. The dashed black line represents the pure elastic response of a tissue with no relaxation due to topological transitions. Inset- visualization the simple shear mode with a given shear rate $\dot{\gamma}$ \label{fig: shear H -0.5 P 1.5}}
\end{figure}

In a constant shear experiment, the tissue is directly sheared along one edge, and the steady state stresses are 
measured as a function of the applied shear rate $\dot \gamma$. The geometry is chosen so that the velocity is in the $x$, shearing the originally perpendicular $y$ direction (see inset in Fig. \ref{fig: shear H -0.5 P 1.5}).   The shear rate defines the metric tensor
\begin{align}
	g_{\mu\nu} = \left( \begin{array}{cc}
		1 & \dot\gamma t \\ \dot\gamma t & 1+ \dot\gamma^2 t^2\
	\end{array}\right),
\end{align}
Given this metric tensor, we study the relaxation of the network tensor driven by the relaxation dynamics 
introduced in  Ref.~\cite{grossman2021instabilities}.
\begin{align}\label{eq: Q_relax}
	\pd_t Q^{\mu\nu} = -\Gamma^{\mu\nu\alpha\beta} \frac{\delta E}{\delta Q^{\alpha \beta}} -\xi_0 Q^{\mu\nu}
\end{align}
where $\eta_{\alpha \beta}=\frac{\delta {E}}{\delta Q^{\alpha \beta}}$ is the conjugate field to the network 
tensor $Q^{\alpha \beta}$, and $\Gamma^{\mu\nu\alpha\beta}= H_1 \left[\frac{1}{2}\left(Q^{\alpha\mu}
Q^{\beta\nu}+Q^{\alpha\nu}Q^{\beta\mu}\right) + H Q^{\alpha\beta}Q^{\mu\nu}\right]$ is the relaxation 
rate tensor. The rate $H_1$ associated to the topological transitions in the tissue  fixes  the time scale 
and we choose in the following the time unit so that $H_1=1$. The dimensionless coefficient $H$ 
{measures} the relative importance of the various topological transitions 
(proliferation versus $T_1$ transitions). For pure $T_1$ transitions that conserve the cell area (no 
proliferation), $H= -\frac{1}{2}$. The last term is due to the active proliferation of the cells
(cell division and cell death), 
with a proliferation rate $\xi_0$. Note that the actual proliferation rate depends on the 
stresses in the system and is therefore different from the bare active proliferation rate $\xi_0$.

Rheological experiments measure stresses. In our model the Cauchy stress is given by
\begin{align}
	\sigma^{\mu\nu} = 
	 \left(\sqrt{g Q} - 1\right) g^{\mu\nu} + 2\frac{k_p}{\sqrt{g Q}} \left(g_{\alpha \beta}Q^{\alpha \beta} - p \right) Q^{\mu\nu}.
\end{align}

As the metric tensor is time dependent, it is convenient to define the time dependent flat-frame, in which 
the metric tensor is the unit tensor. In this frame, the network tensor describes the observable network 
structure in real space. The coordinate transformation uses the elongation tensor $F^\mu_\alpha(t)$, 
which transforms the coordinates in the reference frame into the actual coordinates in real space. It  is 
such that $
(F^{-1})^\mu_\alpha (F^{-1})^\nu_\beta  g_{\mu\nu} = \delta_{\alpha\beta}$. In this frame, the network tensor reads
	\begin{align}
	\label{eq:refframe}
	\Qstar^{\mu\nu} &= F^\mu_\alpha F^\nu_\beta Q^{\alpha \beta}. 
\end{align}
The time derivative of the observable network tensor $\pd_t \Qstar^{\mu\nu}$, is obtained from eq. \eqref{eq: Q_relax}
\begin{eqnarray}\label{eq: Q_flat}
	&\pd_t {\bf\Qstar} =  - \frac{\bf\Qstar}{2\sqrt{\Qstar}}
	\bigl[(1+2H)[(\Qstar-1)-k_p (\Tr {\bf\Qstar}-p)^2] \nonumber\\
	&+4k_p(\Tr {\bf\Qstar}-p)({\bf\Qstar}+H Tr {\bf\Qstar}) \bigr] -\xi_0 {\bf\Qstar}
	+ \bf\lambda,
\end{eqnarray}

The last term is the driving {force of the deformation} due to the imposed shear: 
$\lambda^{\mu\nu} = \pd_t F^\mu_{\alpha}(F^{-1}) ^\alpha_\gamma {Q^*}^{\gamma\nu} +
{Q^*}^{\mu\gamma} (F^{-1}) ^\alpha_\gamma \pd_t F ^\nu_{\alpha}$. It is given explicitly by
\begin{align}
	{\bm \lambda} = \left(\begin{array}{cc}
		0 & \dot\gamma \Qstar^{xx}  \\ 
		\dot\gamma \Qstar^{xx} & 2 \dot\gamma \Qstar^{xy} 
	\end{array}\right)
\end{align}

The determinant of the observable network tensor gives the area of the cells. It is obtained from the following dynamical equation, which is independent of the forcing term ${\bf \lambda}$ 
\begin{equation}
\label{determinant}
 \pd_t \Qstar = -\Qstar^{1/2}(1+2H)\left(\Qstar-1+	 k_p (Tr^2 {\bf\Qstar}- p^2)\right) -  2\Qstar \xi_0
\end{equation}

We first study the case where there is no active cell division ($ \xi_0=0$) and the number of cells is constant. This corresponds to $H=-1/2$ and cells can only move past each other {and exchange neighbors by $T_1$ topological transitions}. The determinant of the observable network tensor and the cell 
area are then constant as imposed by \eqref{determinant}: for a soft solid tissue, when $p>2$ the cell area $A_s$ is given by $A_s=
\Qstar^{1/2}=1$,  whereas for a hard solid tissue,  when $p<2$, $A_s=\frac{1+2 k_p p}{1+ 4 k_p}$.

\begin{figure}
	\centering
	\includegraphics[width=\columnwidth]{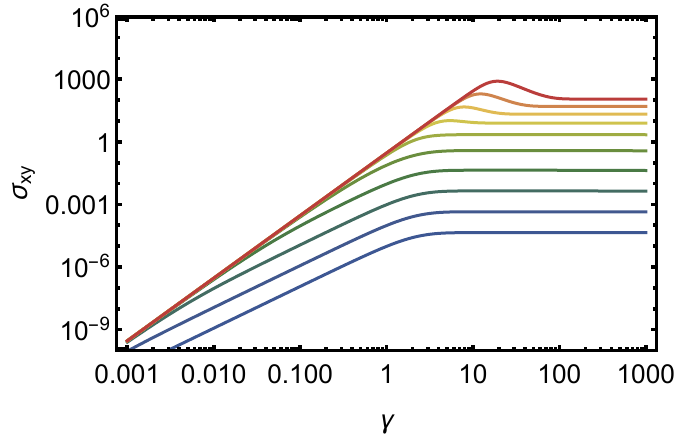}
	\caption{Shear stress as a function of strain, at constant shear rate, when $p=3>2$ and $H= -1/2 $.  The colored  lines correspond to  different shear rates: blue (bottom plot) - $\dot\gamma = 10^{-4}$, red (top plot) - $\dot\gamma = 10^5$. \label{fig: shear H -0.5 P 3}}
\end{figure}

 We solved numerically the dynamical equation for the network 
 tensor \eqref{eq: Q_flat}  and calculated the shear stress as a function of time for a given shear rate. The result 
 is plotted as a function of the strain $\gamma= \dot\gamma t$, in figure \ref{fig: shear H -0.5 P 1.5} for {a} hard solid tissue  ($p<2$) and in figure \ref{fig: shear H -0.5 P 3} for a soft solid tissue ($p>2$). In both cases the tissue is a thixotropic fluid with a time 
 dependent shear thinning behavior. 
For a hard solid tissue, at small strains ($\gamma \ll 1$) corresponding to short times after applications of the shear rate,  the tissue behaves as an elastic solid and shear stress is proportional to strain. 
The elastic shear modulus can be calculated directly from the vertex model in the absence of topological transformations, by expanding the energy to second order {in} strain,


\begin{equation}
	\label{shearmodulus}
	G =2\frac{2-p}{1+4k_p}
\end{equation}

 At larger strains, the stress overshoots and then decays to a constant value at long times
 that only 
depends on shear rate. At low rates, the stress is proportional to  the shear rate defining 
the tissue 
shear viscosity discussed below. At higher shear rate, the tissue is shear thinning. A similar behavior is observed for a soft solid but the shear modulus vanishes and at short times the stress is proportional to the 
square of the strain. The cell orientation measured by the non-diagonal component of the network tensor $Q_{xy}$ is proportional to the shear rate.

At high shear rate $\dot \gamma \gg  H_1$, larger than the rate of $T_1$ transitions, the network tensor 
reaches a steady state with $ \Qstar^{12}=\frac{A_s}{7^{1/2}}$ and $Tr {\bf\Qstar}=(4/7)^{1/6} 
(\frac{A_s^2 \dot \gamma}{k_p})^{1/3}$.  The shear stress can be calculated after performing the base 
transformation 
$\sigma^{xy}=(\frac{4 k_p A_s}{7})^{2/3} \dot\gamma^{1/3}$. The variation of the normal stress difference with strain is given in {appendix D}. At large times it also goes to a constant value $\sigma^{yy}-\sigma^{xx}\sim \dot\gamma^{2/3}$. 

\begin{figure}
	\centering
	\includegraphics[width=\columnwidth]{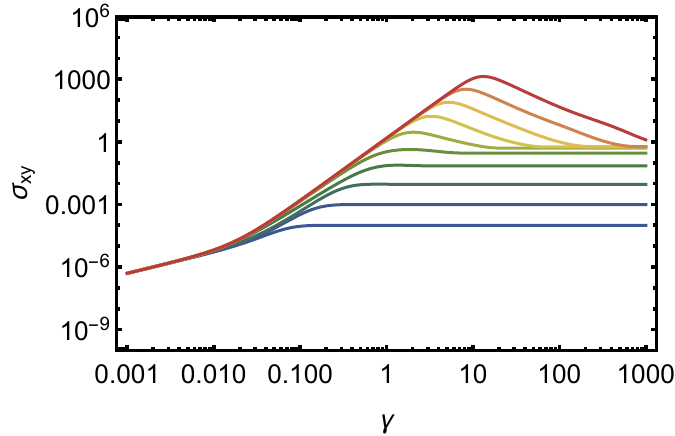}
	\caption{Shear stress as a function of strain, when $p=1.5<2$ and $H=0 $, $\xi_0 = 0.291 < \xi_0^*$.  The colored lines correspond to different shear rates: blue (bottom plot) - $\dot\gamma = 10^{-4}$, red (top plot) - $\dot\gamma = 10^5$. \label{fig: shear H -0 P 1.5 gamma 0.291}}
\end{figure}

\begin{figure}
	\centering
	\includegraphics[width=\columnwidth]{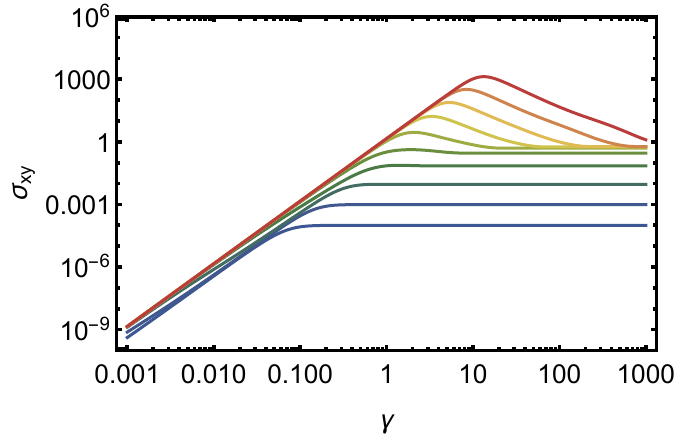}
	\caption{Shear stress as a function of strain, when $p=1.5<2$ and $H=0 $, $\xi_0 = 0.292 > \xi_0^*$.  The colored lines correspond to different shear rates: blue (bottom plot) - $\dot\gamma = 10^{-4}$, red (top plot) - $\dot\gamma = 10^5$.\label{fig: shear H -0 P 1.5 gamma 0.292}}
\end{figure}

In the presence of active cell divisions (when $H> -\frac{1}{2}$ and $\xi_0 \geq 0$)  a somewhat similar pictures arises and is displayed in Fig.~\ref{fig: shear H -0 P 1.5 gamma 0.291} and \ref{fig: shear H -0 P 1.5 gamma 0.292}. The tissue is 
elastic at small deformation or at short times, the stress shows an overshoot at interemediate 
deformations and a plateau at long times. One important result is the existence of a 
shear homeostatic state at long times where the growth rate vanishes and the observable network 
tensor has a constant value. At large shear rates, the long time stress plateau  does not depend 
on the shear rate.  Additionally, for a hard tissue (when $p<2$), there is a dynamical 
transition at the  division rate $\xi_0^* = (1+2H)\frac{4-p^2}{4p}$, which we already discussed in 
Ref~\cite{grossman2021instabilities} for the homeostatic pressure. The mechanical properties of 
the tissue at low strain (or short times) depend on whether the active division rate is smaller or 
larger than this critical value: for $\xi_0<\xi_0^*$ the stress is linear, while for $
\xi_0 > \xi_0^*$ it is {close to} quadratic (except very close to the critical value  $
\xi_0^*$). At high shear rates, the {steady state} stress plateau does not depend on the active division rate  
$\xi_0$. We calculate the high shear rate steady state exponents by power counting as 
above:
$Q =\det \mathbf{Q} \propto \dot{\gamma}^{-1}$, $\Qstar^{12} \propto \dot\gamma^{-1/2}$, and $Tr\mathbf{Q} = const.$ Hence, $\sigma^{xy}= const$, and $\sigma^{yy}-\sigma^{xx} \propto \dot\gamma^{1/2}$.  The exact expressions are given in {appendix D}

\begin{figure}
	\centering
	\includegraphics[width=\columnwidth]{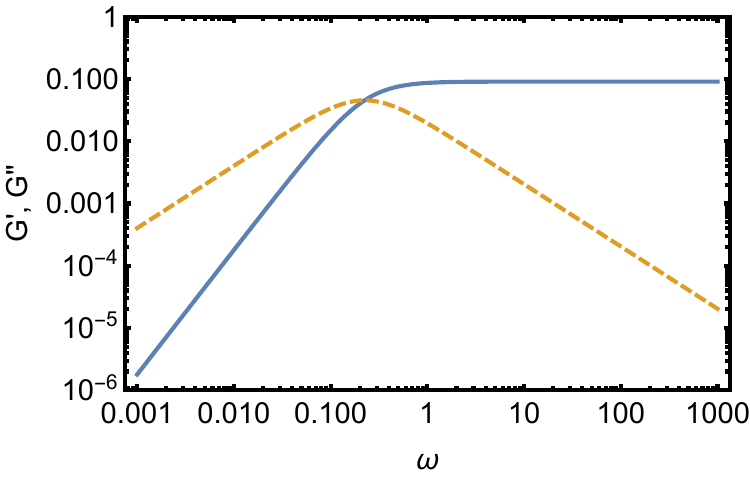}
	\caption{Linear response of a hard tissue $p<2$ under oscillatory forcing, as a function of frequency. This rheology corresponds to that of a viscoelastic Maxwell model. The storage modulus $G'$ is given by the blue lineand  the Loss modulus $G''$ by the dahsed yellow line. \label{fig: maxwell}}
\end{figure}

Using a similar approach, we discuss the linear rheology of the tissue at a finite pulsation $\omega$. In this experiment, a tissue is sheared in an oscillatory manner  so that $\dot \gamma \propto \sin \omega t$.	 A soft 
solid tissue has soft deformation modes and its complex modulus vanishes. The shear aligns the cells in a direction either parallel or perpendicular to the shear. The explicit calculation of 
the complex shear 
modulus of a hard solid tissue  is given on Fig~\ref{fig: maxwell}. It is explicitly calculated by linearizing  {Eq.~} \eqref{eq: Q_flat}, writing $\mathbf{\Qstar} = {\bf  \Qstar}_0 +{\bf \Qstar}_1(\omega) e^{-i \omega t}$, where $$ {\bf  \Qstar}_0 = \left( \begin{array}{cc}{q_0} & 0 \\  0 & q_0
\end{array}\right) $$ is the free isotropic solution as in {Ref.~}\cite{grossman2021instabilities}, and $q_0 = \frac{1+2k_p p}{1+ 4 k_p}$ when proliferation is not allowed ($H=-\frac{1}{2}$), or $q_0 = \sqrt{\frac{1+ k_p p^2}{1+4k_p}}$ when proliferation is allowed ($H>-\frac{1}{2}$).{ Eq.~}\eqref{eq: Q_flat} can then be written as 
\begin{align}\label{eq: linear rheology}
-i \omega {\bf \Qstar}_1= - {\bf  M}{\bf \Qstar}_1 + \bm \lambda_0
\end{align} 
{where} $ \bm \lambda_0$ is the zeroth order expansion of the {term  driving the deformation}. The components of $\mathbf{M}$ are explicitly given in appendix C. Within the linear regime, the tissue behaves as a Maxwell fluid with a high frequency shear 
modulus given by 
\eqref{shearmodulus} and a visoelastic relaxation time
$ T_r = \frac{1}{4k_p q_0- 2k_p p }$.
The tissue viscosity is given by the standard  Maxwell formula $\eta= G T_r$ 

 \section{Tissue on an elastic substrate}
 
\begin{figure}[!h]
	\centering
	\includegraphics{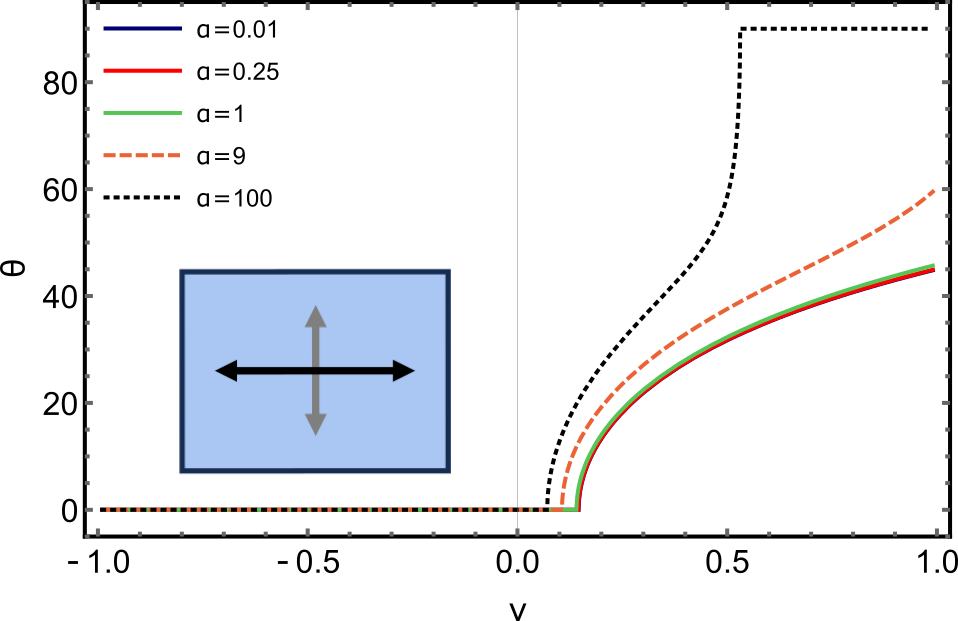}
	\caption{Average tissue angle of compatible tissues supported on an oscillating elastic medium, as a function of the material's Poisson's ratio, for different oscillation amplitudes. Inset on the lower left - visualization of the forcing mode. Black arrow - stretching axis, grey - substrate response (Poisson's ratio), both force the tissue supported by the substrate. \label{fig: oscillation all}}
\end{figure}

We now consider a tissue {forming a} confluent monolayer supported by an elastic 
substrate, with a Poisson ratio $\nu$,  stretched and relaxed in an oscillatory manner along the $x$ 
direction at a pulsation $\omega$ as studied experimentally in {Ref.~}\cite{geremie2022evolution}, {as explained in the} inset of fig. \ref{fig: oscillation all}. We consider that  the tissue strongly coupled to the much stiffer substrate. 
Therefore, the metric tensor in the  tissue is imposed by the elastic deformation of the  substrate. Considering the substrate as a thin elastic sheet, we write it as
\begin{align}\label{eq: time_metric}
	g_{\mu\nu}(t) = \left(\begin{array}{cc}
		1+\alpha \sin^2 \frac{\omega}{2}t  & 0 \\
		0 &(1+\alpha \sin^2 \frac{\omega}{2}t)^{- \nu}
	\end{array}\right)
\end{align}
where $\alpha$ is the stretching amplitude. 

At the time scales of the experiment in Ref.~\cite{geremie2022evolution}, we consider here that there is no cell division or cell death so that the total number of cells $N_{cell} = 
\int \frac{1}{\sqrt{g Q}} \sqrt{g} \dt^2 x$ remains constant. The reason is two fold: first, in what follows, this allows for an easy decoupling of the tissue's mechanical response relative to the substrate, from the effects of proliferation. Second, this is often the case, as cell division rate is typically tens of minutes to many hours, while the oscillation can be significantly faster \cite{geremie2022evolution}.  The deformations of the cells are driven by  the deformation of the substrate 
but  they do not follow it exactly: over long time scales,  cells can change their orientations and their areas from the values 
imposed by the substrate  by detachments from and reattachments to the substrate, which change the 
network tensor ${\bf Q}$. {Detachments and reattachments therefore play the same role as the topological transitions in the previous section.} In a mean field approximation where all cells are identical, the energy given by 
Eq.~\eqref{eq: avg energy} can be written as 
\begin{align}\label{eq: mean energy}
	E(t) = & N_{cells} \Bigg\{\left[\left(1+\alpha \sin^2 \frac{\omega}{2}t\right)^{\frac{1-\nu}{2}}\sqrt{Q(t)} - 1\right]^2   \\ \nonumber &+ k_p \left[\left(1+\alpha \sin^2 \frac{\omega}{2}t\right)^2 Q^{xx}(t)  \right. \\ \nonumber  &\left.+ \left(1+\alpha \sin^2 \frac{\omega}{2}t\right)^{-\nu}Q^{yy}(t) - p\right]^2\Bigg\}  \\ \nonumber
\end{align}

Since the metric tensor is imposed by the substrate, the network tensor, $Q^{\mu\nu}$ is {can}
relax. The symmetries being the same, we use the relaxation dynamics of \eqref{eq: Q_relax} in the 
absence of cell division ($\xi_0=0$). In this experiment, the dynamics {neighbor exchange} ($T_1$ transitions) but not other topological transitions (proliferation). They are {due to}  
attachments and detachments of the cells at a rate $H_1 $ and the 
dimensionless coefficient $H$ that controls the area changes during {these processes}.

If the oscillations are extremely fast,  $\omega \gg H_1=1$, they may be  averaged in the vertex energy. (The explicit calculation is given in appendix A). The average energy reads
\begin{align}\label{eq: E_eff_final}
	\tilde{E}_{eff} = & \frac{2\pi E_{eff}}{N_{Cells}}  =  \Big[ F_{\left((1-\nu),\alpha\right)} Q-2 F_{\left(\frac{1-\nu}{2},\alpha\right)} \sqrt{Q}+1 \Big] \\ \nonumber 
	&+ k_p \Big[ F_{\left(2,\alpha\right)}\left(Q^{xx}\right)^2  + F_{\left(-2\nu, \alpha\right)}\left(Q^{yy}\right)^2   + \left(p\right)^2 \\ \nonumber 
	& +2 F_{\left(1-\nu	,\alpha\right)} Q^{xx}Q^{yy} -2 p \left(F_{\left(1,\alpha\right)} Q^{xx} + F_{\left(-\nu,\alpha\right)}Q^{yy}\right)\Big].
\end{align}

where $ F_{\left(\beta,\alpha\right)} = {}_{2} F_1\left(\frac{1}{2},-\beta,1,-\alpha\right) = 	\int_{-1}^1 \frac{\left(1+ \alpha x^2\right)^\beta}{\sqrt{1-x^2}} \dt x$ is the ordinary hyper-geometric function \cite{bailey1935generalized}.

After a long time, we assume that the system reaches a steady state and we look for a network tensor $Q_0^{\mu\nu}$ that minimizes the energy. A complete derivation of the steady state network tensor is given in appendix B.

In {Eq.~}\eqref{eq: E_eff_final}, the non-diagonal component of the network tensor $Q^{xy}$ appears 
only in the cell area term through  the 
determinant $Q$. {It} can be chosen as an order parameter for the cell 
orientation, because if it does not vanish,  the network is not oriented along the principal axes of the 
deformation. In the following, we focus on $Q^{xy}$ and on the orientation angle of the eigen-direction 
of the network tensor $Q^{\mu\nu}$.

For any oscillation amplitude $\alpha$, and Poisson  ratio $\nu$, $Q^{yy} \geq Q^{xx}$, the component of the network 
tensor perpendicular to deformation axis is larger than the component parallel to the deformation and we  
measure the orientation angle with respect to the $y$ axis. This prefered orientation can be understood 
from an intuitive argument,  that replaces a cell by a thin rod  with finite dimensions. For any Poisson ratio
$-1<\nu<1$, the actual change of the rod's  length  along the $y$ direction is smaller by a factor of $\nu$ 
than along the stretching $x$ direction. It is therefore in general energetically favorable to align along the 
perpendicular direction.

Fig. \ref{fig: oscillation all} shows the orientation angle of the cells in the  case where the shape 
parameter is $p=3$, corresponding to a "soft" solid behavior.  In the absence of any oscillations, 
the tissue is in an ordered state where all the cells are elongated such that {the cell area is} $A=1$ and {$p= 
p^0$}. There is a well defined  cell orientation which is spontaneously chosen. The figure shows the average angle of the cells with respect to the direction perpendicular to the stretching during the oscillations.

Several features emerge from the figure. While for $\nu <0$, $Q^{12}=0$ and the cells align 
perpendicular to the stretching direction, for $\nu>0$ there exists a critical value $\nu^*(\alpha,
p)$, so that for $\nu>\nu^*$ the cells orient at a finite angle $\theta$  relative to the direction 
perpendicular to the stretching. At very large deformations $\alpha \gg 1$, the actual average angle 
during oscillations may reach $\frac{\pi}{2}$ i.e - along the stretching direction. This seems contrary 
to  the intuitive argument given above, but for large 
deformations, the average change of the length of the cell  along the stretching direction
may be arbitrary large (growing as $\alpha$), while in the perpendicular direction its length vanishes 
(roughly as $\alpha ^{-2\nu}$), so that any finite shape is inevitably extremely elongated.

To understand the existence of  finite angle for $\nu>\nu^*$, it is sufficient  again to consider the 
case of a finite length rod. In this case, when $\nu>0$, as the $y$ direction is being stretched, the $x$ direction contracts, therefore there exists an orientation, $\psi$, along which the rod's  length does not change. Rather, at this orientation, it rotates. 
For a thin rod this angle is given by  $ \psi = \text{arccot}\left[\sqrt{\alpha}/{\sqrt{2 -2 F_{(-{\nu},
\alpha)}}}\right]$ (up to reflections around the principal axes).  For a finite cell the range of existence of such an orientation 
depends on $p$. We  expect that as $p \rightarrow \infty$, the onset of a nontrivial orientation 
occurs for  a value of the Poisson ratio $\nu^* \rightarrow 0$, while for $p<2 $ no such solution can be found. Other properties of the tissue are given in {appendix G} including the behavior of the tissue  just after stopping the oscillations

In the limit $\alpha \rightarrow 0$,  we expect that the network tensor $Q^{\mu\nu}$ is close to that of a free standing tissue.  However, the existence of a symmetry breaking transition at a finite Poisson  ratio $\nu^*$, indicates a non perturbative solution. We therefore expand ${\bf Q}(\alpha)$ in the limit $\alpha \rightarrow 0 $ for $p >2$. We call $\tilde {\bf  Q} $ the value of the network tensor that minimizes the energy \eqref{eq: E_eff_final} when $\alpha=0$
	\begin{align}
	\tilde{\bf Q} = \left(\begin{array}{cc}
		\frac{1}{2}\left(p- K \cos 2\theta \right)  & \frac{1}{2}K\sin 2\theta   \\ 
		\frac{1}{2}K\sin 2\theta   & \frac{1}{2}\left(p +K \cos 2\theta  \right)\end{array}\right),
\end{align}
where $K=\sqrt{{p}^2-4}$. Note that the trace is $\Tr \tilde{\bf Q} =p$ and the determinant
$\tilde{Q} =1$. The orientation angle $\theta$ of the 
eigenvectors of the network tensor with respect to the direction perpendicular to the stretching 
direction is unknown at this point. In order to find it, we use the general decomposition of a positive definite network tensor, 
$Q^{\mu\nu}$ in terms of its trace $\bar{q}$ determinant $Q$, and orientation angle 
$\theta$ of its eigenvector
\begin{align}
	Q^{\mu\nu} = \left(\begin{array}{cc}
		\frac{1}{2}\left(\bar{q} - K' \cos 2\theta \right) & \frac{1}{2} K' \sin 2\theta   \\ 
		\frac{1}{2}K' \sin 2\theta    & \frac{1}{2} \left(\bar{q} + K' \cos 2\theta  \right)\end{array}\right).
\end{align}	
where $K'=  \sqrt{\bar{q}^2-4 Q}$.

We expand $\bar{q}=p +\alpha \bar{q}_1 + \alpha^2 \bar{q}_2$, $Q =1+\alpha Q_1 + \alpha^2 Q_2$, and the network tensor, which minimizes the energy \eqref{eq: E_eff_final} up to second order in the amplitude $\alpha$. The minimization of the energy gives an angle $\theta$ independent of $\alpha$. There is therefore  a non-analytical symmetry breaking solution driven by the perturbation due to the stretching.
\begin{align}\label{eq: angle}
	\theta =& \left\{ \begin{array}{c c}
		\frac{\pi}{2} \pm \frac{1}{2}\arccos\left(-\frac{p}{\sqrt{p^2-4}}\frac{1-\nu}{1+\nu}\right)  &,~  \nu \geq \nu^* \\ 
		0 &,~ \nu<\nu^* \end{array} \right.
\end{align}
where the sign corresponds to the remaining reflection symmetry,  and $\nu^* = \frac{p -\sqrt{p^2-4}}{p +\sqrt{p^2-4}}$,
For any $\nu > \nu^*$, $\theta \neq 0$ and $\theta (\nu =1)=\pi/4$.

When $p<2 $, in the limit of small amplitude $\alpha \rightarrow 0$, the network tensor is isotropic
\begin{align}
	\tilde{Q}^{\mu\nu} = \left(\begin{array}{cc}
		\frac{1+2 k_p p}{1+ 4 k_p} &0 \\ 
		0 & \frac{1+2p  k_p }{1+ 4 k_p }\end{array}\right).
\end{align} 

For a tissue on a stretched substrate, the rheology at finite pulsation $\omega$ can be discussed as in the previous section.
We use  the same procedure as for the shear geometry  and perform the change of reference 
frame to the real coordinates given by \eqref{eq:refframe}. The relaxation dynamics of the network 
tensor is given  by \eqref{eq: Q_relax} in the absence of cell divisions ($\xi_0=0$). 
Note however that the field $\bm \eta= \frac{\delta E} {\delta \bf Q}$ conjugate to the network 
tensor 
must be calculated at constant number of cells. This is done by introducing the energy per cell $e$ 
such that $E=N_{cell} \cal{E}$. The conjugate field is then calculated as $\bm \eta=\rho \frac{\partial \cal{E}} 
{\partial \bf Q}$.
The relaxation dynamics of the observable network tensor in given by 

\begin{align}\label{eq: Q_flat2}
	\pd_t {\bf \Qstar} = & - \sqrt{\Qstar}\left(\sqrt{\Qstar}-1\right)\left(1+2H\right){\bf \Qstar} \\
	\nonumber 
	& - 2k_p {\bf \Qstar} \left(\Tr {\bf \Qstar} - p \right) \left({\bf  \Qstar}+ H \Tr { \bf \Qstar}  \right) \\ \nonumber
	&+{\bm \lambda},
\end{align}
where the driving term is 
\begin{align}
 {\bm \lambda} = \left(\begin{array}{cc}
		2 f\left(\omega,t,\alpha\right) \Qstar^{11} & (1-\nu)f\left(\omega,t,\alpha\right) \Qstar^{12}  \\
		(1-\nu)f\left(\omega,t,\alpha\right) \Qstar^{12} & - 2\nu f\left(\omega,t,\alpha\right) \Qstar^{22}
	\end{array}\right)
\end{align}
with a driving function $f\left(\omega,t,\alpha\right) = \frac{\alpha {\omega}}{4} \frac{\sin  \omega t}{1+ \alpha \sin^2 \frac{\omega}{2} t}$ 

The geometry induces a highly non-linear driving term and we only study here the properties of the tissue at  linear order in the amplitude of the oscillation. The linear response is given by \eqref{eq: linear rheology}. Here the small parameter is the amplitude $\alpha$, and we use complex notations replacing $\sin(\omega t)$ with the complex exponent $e^{-i \omega t}$.

For a tissue on  a substrate, the stress response of the tissue is less important experimentally as the measured stress would correspond to the sum of the substrate stress and the tissue stress ${\bf \sigma}_1$, and by assumption, the substrate is much stiffer than the tissue.  We give the calculated value of the tissue stress in {appendix F}. The results of the hard tissue are similar to those in the previous section and the tissue behaves as a Maxwell visco-elastic fluid with the modulus and relaxation time obtained in the previous section.

In order to calculate the finite frequency response of a soft tissue, we expanded {the network tensor} $\bf Q$ around the ground state with a definite, yet still unknown{:} $\mathbf{\Qstar} = \tilde{\bf Q}(\theta) + {\bf \Qstar}_1$. {We then solved} for ${\bf \Qstar}_1$ and {calculated} the angle that minimizes the energy $E(\tilde{\bf Q}(\theta) + {\bf \Qstar}_1)$ expanded to second order in the amplitude $\alpha$. {We anticipated here that the periodic forcing leads  at very long times to a steady state static solution, breaking the rotational symmetry}.  

\begin{figure}
	\centering
	\includegraphics[width=\columnwidth]{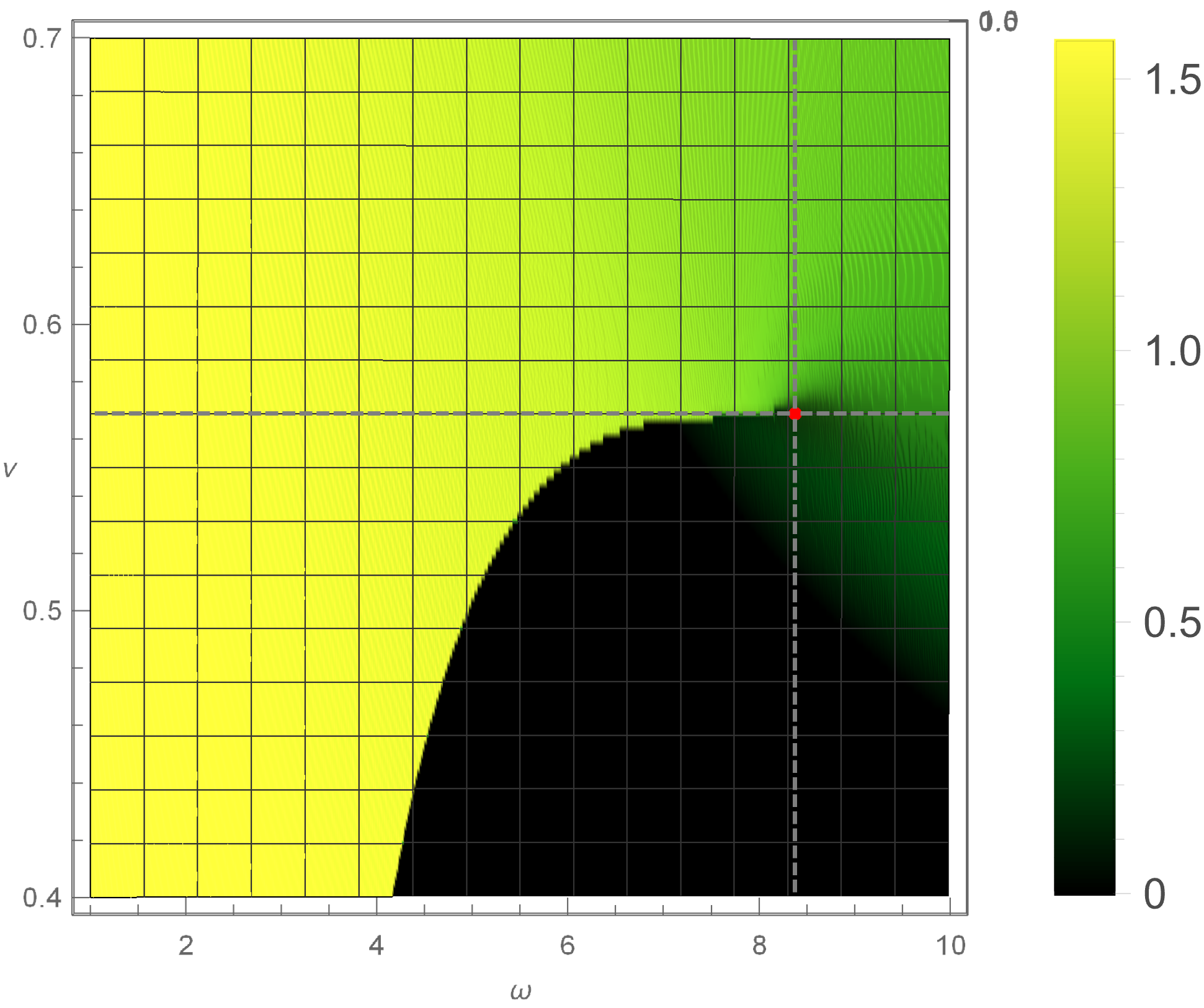}
	\caption{Tricritical behavior of the cells average orientation $\theta(\omega,\nu)$, as a function of frequency, $\omega$, and Poisson's ratio, $\nu$. The colors corresponds to the values of the angle $\theta$: dark green - $\theta =0$, bright yellow = $\theta = \pi/2$. Here $k_p =1$, $p=3$ $H=0$. Red dot marks the tricritical point ($\omega_c \sim 8.5$  $\nu_c \sim 0.53$) Above $\omega_c$ the transition is continuous, as indicated by the smooth color gradient. Below  the critical  value, the transition is discontinuous. $\theta =0 $ corresponds to cell oriented perpendicular to the main stretching direction, $\theta =\pi/2$ corresponds to a parallel orientation to the stretching.\label{fig: tricrit} }
\end{figure}

The resulting phase space $\theta(\omega,\nu)$ is quite complicated, and  exhibits a tricritical point. {We show in  Fig. \ref{fig: tricrit} the detailed  phase diagram} around the tricritical point $\omega_c\sim 8.5$, and $\nu_c \sim 0.58$. {More details on the phase  space and the tricritical point are given in appendix H}.

\subsection{Discussion}
This paper discusses the macroscopic mechanical properties of an epithelial two dimensional tissue starting from 
the cellular vertex model introduced in reference \cite{grossman2021instabilities}. As emphasized 
in this work, the rheology of the tissue strongly depends on the existence of 
internal stresses in the tissue (hard solid for a shape factor $p<2$) or on soft modes (soft solids 
for a shape factor $p>2$).  We study explicitly two geometries, a sheared tissue and a tissue 
on a stretched elastic substrate.

{The main result of our work is an explicit analytical calcualtion of the rheological properties of a tissue, using a coarse-grained model that is rigorously derived from a detailed, discrete, vertex model.} 
An important general result is that the tissues can be considered as thixotropic fluids. After the application of a constant shear rate the tissues are elastic at short time with a finite shear modulus for hard tissues and a vanishing shear modulus for soft tissues. As the strain is increased, the stress goes through a maximum and then decreases to a plateau that depends only of the applied shear rate. At small shear rates the tissue is liquid with a finite viscosity. At large shear rates, the tissue is shear thinning and the stress increases as  a power low of the shear rate. As in previous numerical studies \cite{matoz2017cell, basan2011dissipative}, we never find a finite yield stress even in the absence of cell proliferation. The stress always 
relaxes because of the cell rearrangements due to the $T_1$ topological processes (or due to proliferation if present). A shear thinning behavior has already been observed in numerical particle simulations, or predicted in simplified models for tissues in Ref.~\cite{basan2011dissipative,ishihara2017cells}. 

An important result for proliferating tissues is the existence of a shear homeostatic state which is a steady state of the tissue under shear where cell division  exactly compensates cell death. In this steady state the tissue has plastic behavior where at long times the stress reaches a plateau independent of shear rate and of the active division rate.  

We also discussed the linear rheological behavior at finite frequency of hard tissues. They behave as Maxwell fluids with a finite relaxation time due to the topological transformations in the tissue in line with previous work on cell division \cite{ranft2010fluidization}. Because of the presence of soft modes, the linear complex modulus of soft tissues vanishes.

This general picture of two-dimensional tissue rheology seems to be robust as attested by the existing numerical simulations of tissues. Our results also share some similarities with the rheology of foams \cite{khan1988foam,kraynik1988foam,sibree1934viscosity}.  Still some details of the predictions such as the shear thinning exponent $1/3$ might depend on the precise assumptions made for the transport coefficient $\Gamma^{\mu\nu\alpha\beta}$ in {Eq.~}\eqref{eq: Q_relax}. We chose the simplest transport coefficient compatible with the symmetries and that depends only on the network tensor {which is the tensor that we use to characterize the topological transitions in the tissue}. Contributions of the metric tensor and more general non-linear terms in order to go beyond the geometric non-linearities could be introduced and modify the shear thinning exponent. We also have considered that the metric tensor measuring  elastic effects relaxes intantaneously, which is certainly not true at high frequencies. The elastic relaxation could be taken into account by the introduction of a short time viscosity.

The second geometry that we considered is that of experiments measuring cell orientation under periodic stretch of an elastic substrate. Only elongated cells orient under stretch, corresponding to soft tissues. If the stretching is performed at high frequencies, our results at small amplitudes are very similar to those obtained from a macroscopic elastic description of the tissue as a uniaxial material \cite{livne2014cell,moriel2022cellular, geremie2022evolution}. We provide here extensions of these results to non-linear stretching amplitudes and to finite frequencies.  

Finally, we only considered homogeneous uniform tissues. Adding seams or non-uniform regions where the tissue might be stiffer (simulating scars), or just has a different topology (such as a hole) can shed important light on the way tissues heal and help better understand how scars  affect the mechanical response of an organ. 



\bibliography{D:/Users/grossman/Documents/PostDoc_Old/Projects/Papers_and_Notes/bibs/active}

\end{document}


\title{Predicting the mechanical properties of spring networks - Appendixes}
\date{\today}

\author{Doron \surname{Grossman}}
\email[]{doron.grossman@ladhyx.polytechnique.fr}
\affiliation{College de France, 11 Pl. Marcelin Berthelot, 75231 Paris}
\author{Jean-Francois \surname{Joanny}}
\affiliation{College de France, 11 Pl. Marcelin Berthelot, 75231 Paris}
\date{\today}

%
%
%

\keywords{epithelial cells $|$ rheology $|$ tissue mechanics $|$ out of equilibrium} 

\maketitle
	\section{Appendix A - Derivation of the energy for fast oscillation energy}
	In this appendix we derive the effective average energy of a tissue supported on a fast oscillating substrate.

	We integrate the energy over time
	\begin{align}\label{eq: E_eff(t)}
		\mathcal{E}_{eff}(t) =& \frac{\int_0^T \mathcal{E}(t) \dt t}{\int_0^T \dt t} \\ \nonumber =& \frac{1}{4\pi} \int_{-1}^{1} \Big[  \left((1+\alpha x^2)^{\frac{1-\nu}{2}}\sqrt{Q^{11}(t)Q^{22}(t)-\left[Q^{12}(t)\right]^2} - 1\right)^2   \\ &+ k_P \left(\left(1+\alpha x^2\right) Q^{11}(t) + \left(1+\alpha x^2\right)^{-\nu}Q^{22}(t) - \ptwo^0\right)^2 \Big] \frac{\dt x}{\sqrt{1-x^2}},
	\end{align}
	where we defined $x=\sin\frac{\omega}{2} t$.
	Using
	\begin{align}
		\frac{1}{\pi}\int_{-1}^1 \frac{\left(1+ \alpha x^2\right)^\beta}{\sqrt{1-x^2}} \dt x =F_{\left(\beta,\alpha\right)} = {}_{2} F_1\left(\frac{1}{2},-\beta,1,-\alpha	\right),
	\end{align}
	where ${}_2F_1\left(a,b,c,z\right)$ is the ordinary hyper-geometric function defined in \cite{bailey1935generalized,weisstein2002hypergeometric}, we obtain the result given in the main text. 
	
	\section{Appendix B - Steady state network tensor for fast oscillations}
	Here we give exact expressions for the network tensor of a tissue supported on a fast oscillating substrate.
	
	We write the network tensor minimizing the energy in the limit of fast oscillations using the trace $q= {Q^*}^{11} + {Q^*}^{22}$ and the determinant $Q= {Q^*}^{11}{Q^*}^{22}-({Q^*}^{33})^2$:
	\begin{align}
		{Q^*}^{\mu\nu}= \left(\begin{array}{cc}
			\frac{1}{2}\left(q + \sqrt{q^2-4 Q}\cos 2\theta\right) & \frac{1}{2}\sqrt{q^2-4 Q}\sin 2\theta \\\frac{1}{2} \sqrt{q^2-4 Q}\sin 2\theta & 	\frac{1}{2}\left(q - \sqrt{q^2-4 Q}\cos 2\theta\right)
		\end{array}\right)
	\end{align}
	
	The minimization leads to: 
	\begin{align}
		Q = \frac{\fone}{\ftwo}
	\end{align}
	\begin{align}
		q = \ptwo^0 \frac{\ftwo-\fthree + \ftwo \fthree - \ffour + \frac{\alpha}{2}\left(\ftwo-\fthree(2+\frac{3\alpha}{4})-\ffour\right)}{\left(\ftwo\right)^2-\ffour\left(1+\alpha+\frac{3\alpha^2}{8}\right)}
	\end{align}
Up to reflections about the principal axes of the strain, the orientation angle of the cells reads:
	\begin{align}
		\cos 2\theta =\frac{ \ftwo \ptwo^0 \left(\Xi - 4 \ffour\left(2+\alpha\right) \right)}{\sqrt{\left(\ftwo \ptwo^0\right)^2 \left[\Xi + 4 \ffour\left(2+\alpha\right) \right]^2- 4 \left(\fone\right)^2\left[\ffour(8+8\alpha+3\alpha^2) - 8 \left(\ftwo\right)^2\right]^2}}
	\end{align}
	where  $\Xi =4 \alpha (2 \fthree-\ftwo)+ 8\ftwo (\fthree-1)+8\fthree+3 \alpha^2 \fthree$

	\section{Appendix C -Linear relaxation matrix M of a free tissue at finite frequency} 
	Here we supply exact expressions for the response matrix at  the linear regime.
	
	We use the notation   ${\bf M'}={\bf M }- 2 i \omega {\bf I}$, where $\bf I$ is the identity matrix. and calculate the matrix {\bf M' }
	\begin{subequations}
		\begin{align}\nonumber
			M'^{11} =& -\frac{1}{{2 \Qstar_0}}\left\{k_P \Qstar_0 \left[-\left(1+H\right)\left(2\ptwo^0 -3\Qstar_0^{11}\right)\Qstar_0^{11}+\left(-H \ptwo^0 +2\left(1+2H\right) \Qstar_0^{11}\right)\Qstar_0^{22} + H \left(\Qstar_0^{22}\right)^2+ \left(\Qstar_0^{12}\right)^2\right] \right. \\ & \left.+\left(1+2H\right)\left(\left(4\Qstar_0 -3\right)\Qstar_0^{11}\Qstar_0^{22}-2\left(\Qstar_0-1\right)\left(\Qstar_0^{12}\right)^2\right) +4 i \omega \Qstar_0\right\}
		\end{align}
		\begin{align}\nonumber
			M'^{22} =&-\frac{1}{{2 \Qstar_0}}\left\{k_P \Qstar_0 \left[-\left(1+H\right)\left(2\ptwo^0 -3\Qstar_0^{22}\right)\Qstar_0^{22}+\left(-H \ptwo^0 +2\left(1+2H\right) \Qstar_0^{22}\right)\Qstar_0^{11} + H \left(\Qstar_0^{11}\right)^2+ \left(\Qstar_0^{12}\right)^2\right] \right. \\ & \left.+\left(1+2H\right)\left(\left(4\Qstar_0 -3\right)\Qstar_0^{22}\Qstar_0^{11}-2\left(\Qstar_0-1\right)\left(\Qstar_0^{12}\right)^2\right) +4 i \omega \Qstar_0\right\}
		\end{align}
		\begin{align}
			M'^{33} =& -\frac{1}{2}\left[4\left(1+H\right)k_P \left(\ptwo^0-\Qstar_0^{11}-\Qstar_0^{22}\right)\left(\Qstar_0^{11}+\Qstar_0^{22}\right) -2\left(1+2H\right)\frac{\left(\Qstar_0-1\right)\Qstar_0^{11}\Qstar_0^{22}+\left(2-3\Qstar_0\right)\left(\Qstar_0^{12}\right)^2}{\Qstar_0}-4 i \omega\right]
		\end{align}
		\begin{align}
			M'^{12} =& \frac{\Qstar_0^{22} \left[4 H k_P \Qstar_0 \left(\ptwo^0 -2 \Qstar_0^{11}\right)- \left(1+2H\right)\left(2\Qstar_0 + 4 k_P \Qstar_0 -1\right)\Qstar_0^{22}\right]- 4 k_P \Qstar_0 \left(\Qstar_0^{12}\right)^2}{2 \Qstar_0}
		\end{align}
		\begin{align}
			M'^{21} =& \frac{\Qstar_0^{11} \left[4 H k_P \Qstar_0 \left(\ptwo^0 -2 \Qstar_0^{22}\right)- \left(1+2H\right)\left(2\Qstar_0 + 4 k_P \Qstar_0 -1\right)\Qstar_0^{11}\right]- 4 k_P \Qstar_0 \left(\Qstar_0^{12}\right)^2}{2 \Qstar_0}
		\end{align}
		\begin{align}
			M'^{13} & \frac{-\left(1+2H\right)\left(2\Qstar_0 -1\right)\Qstar_0^{22}+4\left(1+H\right) k_P\Qstar_0 \left[\ptwo^0-2\left(\Qstar_0^{11}+\Qstar_0^{22}\right)\right]}{2 \Qstar_0}\Qstar_0^{12}
		\end{align}
		\begin{align}
			M'^{23}=& \frac{-\left(1+2H\right)\left(2\Qstar_0 -1\right)\Qstar_0^{11}+4\left(1+H\right) k_P\Qstar_0 \left[\ptwo^0-2\left(\Qstar_0^{11}+\Qstar_0^{22}\right)\right]}{2 \Qstar_0}\Qstar_0^{12}
		\end{align}
		\begin{align}
			M'^{32}=& \frac{\left(1+2H\right)\left(2\Qstar_0 -1\right)\Qstar_0^{22}+4 k_P\Qstar_0 \left[\ptwo^0-\left(\Qstar_0^{11}+\Qstar_0^{22}\right)\right]}{ \Qstar_0}\Qstar_0^{12}
		\end{align}
		\begin{align}
			M'^{31}=& \frac{\left(1+2H\right)\left(2\Qstar_0 -1\right)\Qstar_0^{11}+4 k_P\Qstar_0 \left[\ptwo^0-\left(\Qstar_0^{11}+\Qstar_0^{22}\right)\right]}{ \Qstar_0}\Qstar_0^{12}
		\end{align}
	\end{subequations}
	where $\Qstar_0 = \det \Qstar_0^{\mu\nu}$
	
	For a soft solid tissue, in the limit $\alpha \rightarrow 0$ we know that $\Qstar_0 =1$ and $\Qstar_0^{11}+\Qstar_0^{22}=\ptwo^0$.
	When $\nu< \nu^*$ we also know that $\Qstar_0^{12}=0$ hence the expression of the relaxation matrix  $\bf M'$
	\begin{subequations}
		\begin{align}
			M'^{11} &= -\frac{1}{2}\left(1+2H\right) \Qstar_0^{11}\Qstar_0^{22} - 2k_P\left[\left(1+H\right)\ptwo^0 \Qstar_0^{11} - 1\right]-2 i \omega \\
			M'^{22} &= -\frac{1}{2}\left(1+2H\right) \Qstar_0^{11}\Qstar_0^{22} - 2k_P\left[\left(1+H\right)\ptwo^0 \Qstar_0^{22} - 1\right]-2 i \omega \\
			M'^{33} &= \left(1+2H\right) \left(\Qstar_0^{11}\Qstar_0^{22}-1\right) - 2 i \omega \\
			M'^{12} &=  \frac{1}{2} \left\{-\left(1+2H\right) \left(\Qstar_0^{22}\right)^2 -4k_P \left[\left(1+H\right)\ptwo_0^2 \Qstar_0^{22}-1\right]\right\}\\
			M'^{21} &=  \frac{1}{2} \left\{-\left(1+2H\right) \left(\Qstar_0^{11}\right)^2 -4k_P \left[\left(1+H\right)\ptwo_0^2 \Qstar_0^{11}-1\right]\right\}\\
			M'^{13} &= -\frac{1}{2}\sqrt{\Qstar_0^{11}\Qstar_0^{22}-1}\left[\left(1+2H\right)\Qstar_0^{22}+4\left(1+H\right)k_P \ptwo^0\right]\\
			M'^{23} &= -\frac{1}{2}\sqrt{\Qstar_0^{11}\Qstar_0^{22}-1}\left[\left(1+2H\right)\Qstar_0^{11}+4\left(1+H\right)k_P \ptwo^0\right]\\
			M'^{31}&=\left(1+2H) \Qstar_0^{11}\sqrt{\Qstar_0^{11}\Qstar_0^{22}-1}\right)\\
			M'^{32}&=\left(1+2H) \Qstar_0^{22}\sqrt{\Qstar_0^{11}\Qstar_0^{22}-1}\right)\\
		\end{align}
	\end{subequations}

\section{Appendix D - High shear rate exponents}
In order to find the high shear rate exponents, we write the elements of the network tensor $Q^{\mu\nu}$ as exponents of the shear rate:
\begin{align}
	{\bf Q} = \left(\begin{array}{cc}
		\dot{\gamma}^a &		\dot{\gamma}^c\\
		\dot{\gamma}^c & \dot{\gamma}^b
		\end{array}\right).
\end{align} 
Thus $\Tr{\bf Q} =\dot{\gamma}^d$ with $d = max(a,b)$, and $Q=\dot{\gamma}^e=\dot{\gamma}^{a+b} - \dot{\gamma}^{2c}$ so that $e= max(a+b,2c)$. Since the shear is in the $y$ direction and the velocity in the $x$ direction, it is natural to expect  that $b\geq a$. 
Rewriting eq(7) and (9) in the main text, in terms of  $Q$, $\Tr{\bf Q}$ and $Q^{12}$, we obtain, in the limit of  high shear rate $\dot{\gamma}$, the asymptotic behavior of the equation (where prefactors are of no importance):
\begin{align}
	\pd_t Q &\sim \left(1+2H\right)\left(\dot{\gamma}^{\frac{3e}{2}}+\dot{\gamma}^{\frac{e}{2}} + \dot{\gamma}^{2d+\frac{e}{2}}\right) \\ 
	\pd \Tr{\bf Q} & \sim \dot\gamma^{\frac{e}{2}} +\dot\gamma^{d+\frac{e}{2}} + \dot\gamma^{2d+\frac{e}{2}}+ \dot\gamma^{d-\frac{e}{2}} + \dot\gamma^{2d-\frac{e}{2}} + \dot\gamma^{3d-\frac{e}{2}} + \dot\gamma^{1+c} \\
	\pd Q^{12} & \sim \dot\gamma^{\frac{e}{2}+c} +\dot\gamma^{d+\frac{e}{2} + c} + \dot\gamma^{2d+\frac{e}{2}+c}+ \dot\gamma^{c-\frac{e}{2}} + \dot\gamma^{c+ d-\frac{e}{2}} + \dot\gamma^{c+2d-\frac{e}{2}} + \dot\gamma^{1+a}	
\end{align}
The factor $(1+2H)$ was included since it vanishes when  $H=-\frac{1}{2}$ implying  that the  exponent $e=0$.

We divide the problem into two cases: $H= -\frac{1}{2}$ ( no division) and otherwise. In the first case $e=0$, by virtue of the fact that $\pd_t Q = 0 $ at every moment. Assuming $c>0$ and that $d = b$, one finds from eq. (12) of the main text, $3d =  1+c $ and from eq. (13) $2d+c = 1+a$. The value of the exponents is then $ d= \frac{2+a}{5}$ and $c=\frac{1}{5}\left(1+3a\right)$. Requiring that either $e=0=d+a$ or $e=0=2c$  leads to the same result $a=-\frac{1}{3}$,$ z=0$ and $d=\frac{1}{3}$. 

Similarly, when $H>-\frac{1}{2}$, a steady state can occur when either $e\geq 0$ in which case one must require that the additional terms compensate each other, or when  $e<0$, leading to $2d < -\frac{e}{2}$. In fact, only this last case gives consistent results analytically (and is the only one observed numerically).  As before, assuming $d=b$,  it can be shown that $d>0$ is inconsistent, and thus, guessing $d=0$ we are left  with the equations $-\frac{e}{2} = 1+c $ and $c-\frac{e}{2} = 1+a$, leading to $ e= -2 -a$ and $c=\frac{a}{2}$ then either requiring that $b+a =e $ or  $2c=e$ results with $a=-1$ and hence $e= -1$, $ c= -\frac{1}{2}$. Finding the prefactors requires plugging in the asymptotic dependence and comparing coefficients.

\section{Appendix E - Stress in a sehared tissue}

\begin{figure}[!h]
	\centering
	\includegraphics[width=\textwidth]{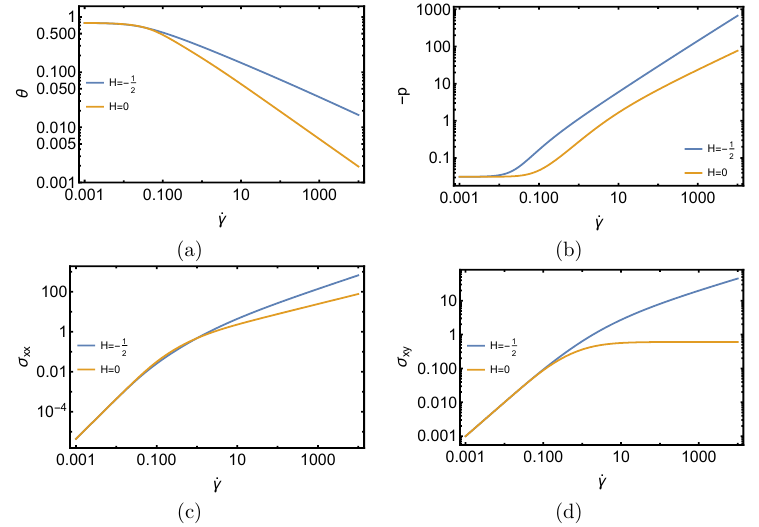}
	\caption{Stress in a hard solid tissue under simple shear$\ptwo^0/A^0=1.5 <2$, $\theta$ is the angle relative to the shear direction, $p=\frac{1}{2}\Tr(\tau)$ is the pressure (it is negative, indicating tension, $\sigma_{xx}$ is the normal stress component, and $\sigma_{xy}$ is the shear stress component.}
\end{figure}
\begin{figure}[!h]
	\centering
	\includegraphics[width=\textwidth]{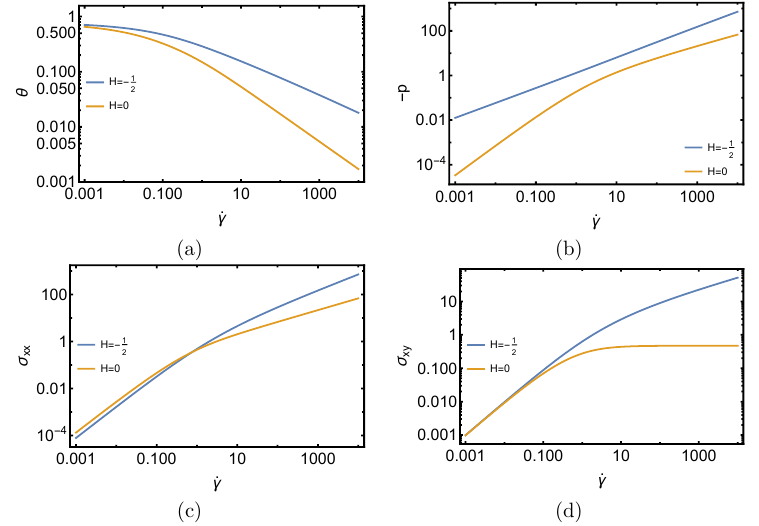}
	\caption{Stress in a marginal tissue under simple shear $\ptwo^0/A^0=2$, $\theta$ is the angle relative to the shear direction, $p=\frac{1}{2}\Tr(\tau)$ is the pressure (it is negative, indicating tension, $\sigma_{xx}$ is the normal stress component, and $\sigma_{xy}$ is the shear stress component.}
\end{figure}
\begin{figure}[!h]
	\centering
	\includegraphics[width=\textwidth]{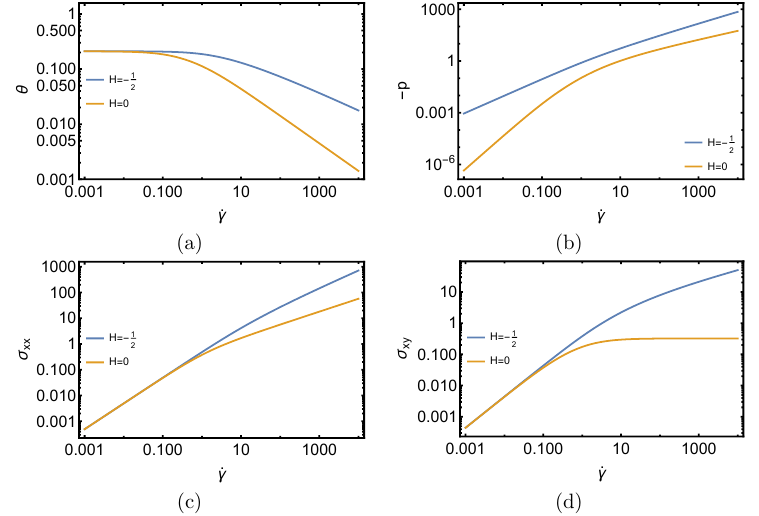}
	\caption{Stress in a soft solid tissue under simple shear $\ptwo^0/A^0=3$, $\theta$ is the angle relative to the shear direction, $p=\frac{1}{2}\Tr(\tau)$ is the pressure (it is negative, indicating tension, $\sigma_{xx}$ is the normal stress component, and $\sigma_{xy}$ is the shear stress component.}
\end{figure}
\begin{figure}[!h]
	\centering
	\includegraphics[width=\textwidth]{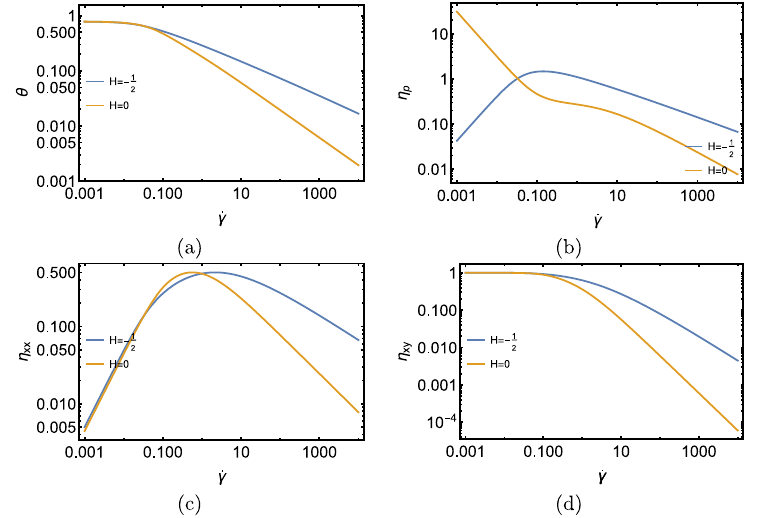}
	\caption{Non-linear viscosity $\eta = \sigma/\dot{\gamma}$ of a hard solid tissue. $\ptwo^0/A^0=1.5$, $\theta$ is the angle relative to the shear direction, $p=\frac{1}{2}\Tr(\tau)$ is the pressure (it is negative, indicating tension, $\sigma_{xx}$ is the normal stress component, and $\sigma_{xy}$ is the shear stress component.}
\end{figure}
\begin{figure}[!h]
	\centering
	\includegraphics[width=\textwidth]{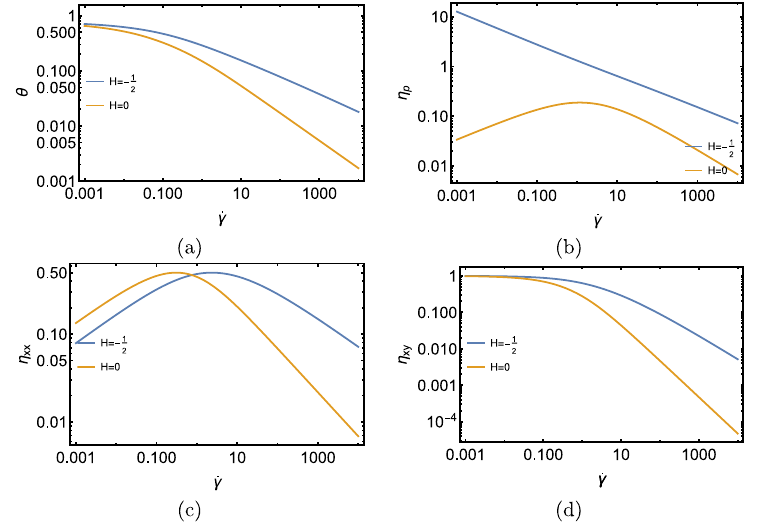}
	\caption{Non linear viscosity  $\eta = \sigma/\dot{\gamma}$ of a marginal tissue.$\ptwo^0/A^0=2$, $\theta$ is the angle relative to the shear direction, $p=\frac{1}{2}\Tr(\tau)$ is the pressure (it is negative, indicating tension, $\sigma_{xx}$ is the normal stress component, and $\sigma_{xy}$ is the shear stress component.}
\end{figure}
\begin{figure}[!h]
	\centering
	\includegraphics[width=\textwidth]{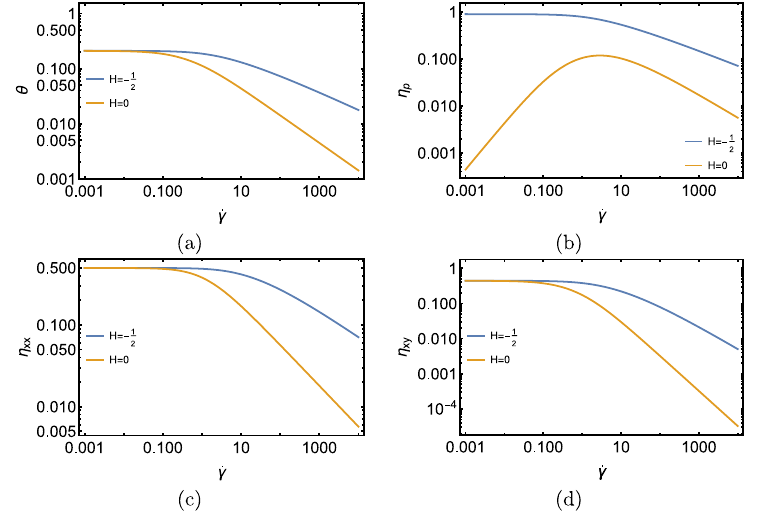}
	\caption{Non-linear viscosity  $\eta = \sigma/\dot{\gamma}$ of a soft solid tissue. $\ptwo^0/A^0=3$, $\theta$ is the angle relative to the shear direction, $p=\frac{1}{2}\Tr(\tau)$ is the pressure (it is negative, indicating tension, $\sigma_{xx}$ is the normal stress component, and $\sigma_{xy}$ is the shear stress component.}
\end{figure}

\clearpage
\section{Appendix F - Linear rheology of a tissue }
Here we write the exact  expressions for  the network tensor for a supported tissue on an oscillating substrate at a finite frequency.

One can expand the stress $\tau^{\mu\nu}$ in powers of the correction to the network tensor ${Q_1^*}^{\mu\nu}$.  The (complex) first order term is ${\tau_1^*}^{\mu\nu}  e^{i \omega t}$, where $\tau_1^*$ is the complex stress. In the case of a tissue on a substrate, where the number of cells is constant,  if $\ptwo < 2$, the complex stress is, to linear order, given by
\begin{align}
	{\tau_1^*}^{\mu\nu} = & \left[\frac{1}{2}\left(2\sqrt{\tilde{Q}}-1\right){\tilde{Q}^{-1}}_{\alpha\beta}{Q_1^*}^{\alpha\beta} 
	+ \frac{2 k_P}{\sqrt{\tilde{Q}}} {Q_1^*}^\alpha_\alpha \right]  \delta^{\mu\nu} 
	\\ \nonumber
	&+ \frac{2 k_P}{\sqrt{\tilde{Q}}}\left(\tilde{Q}^{\alpha}_\alpha -\ptwo\right){Q_1^*}^{\mu\nu}.
\end{align}
Where $\tilde{Q}= \det \tilde{Q}$ is the determinant of $\tilde{Q}^{\mu\nu}$ , $\tilde{Q}^{-1}_{\mu\nu}$ is the inverse of $\tilde{Q}^{\mu\nu}$

When $\ptwo \geq 2$, the complex stress reads
\begin{align}
	{\tau_1^*}^{\mu\nu} = & \frac{1}{2}{\tilde{Q}^{-1}}_{\alpha\beta}{Q_1^*}^{\alpha\beta} \delta^{\mu\nu} + 2 k_P {Q_1^*}^\alpha_\alpha \tilde{Q}^{\mu\nu},
\end{align}
where we used the fact that $\tilde{Q} =1$. The important part here is to note the very different dependence of ${\tau_1^*}^{\mu\nu}$ on ${Q^*_1}^{ \mu\nu}$. Specifically note that any of the shear components of ${\tau_1^*}^{\mu\nu}$ (the traceless part of ${\tau_1^*}^{\mu\nu}$) do not depend on the orientation of ${Q_1^*}^{\mu\nu}$ when $\ptwo \geq 2$. This stems directly from the degeneracy of the energy, relative to an orientation of the cells.

The shear dynamic modulus $G_s = G'_s + i G''_s$ is defined as $G_S=\frac{{\tau_1^*}^{12}}{{\varepsilon^*}}$ where ${\varepsilon^*}$ is the linear complex strain, enforced on the system, defined as ${\varepsilon^*}_{\mu\nu} = F_\mu^ \alpha F_\nu^ \beta\left(g_{\alpha\beta}(t)-g_{\alpha \beta}(0)\right) e^{-i \omega t}$. Similarly, the normal dynamic modulus  $G_n = G'_n + i G''_n = \frac{{\tau_1^*}_n}{{\varepsilon^*}}$, where ${\tau_1^*}_n={\tau_1^*}^{11}-{\tau_1^*}^{22}$ is the normal stress difference
(In this case, $\varepsilon^*_n =\left(\varepsilon^*_{11}- \varepsilon^*_{22}\right) =1+\nu$).

The direct calculation shows that for a tissue on a substrate, $G_n$  when $p<2$, is exactly the same as the shear modulus in the main text when $H=-\frac{1}{2}$. i.e, the system behaves as a Maxwell material. Additionally, note that as $\ptwo$ approaches the critical value $\ptwo \rightarrow 2$, $G_n \rightarrow 0$, stemming from the fact that this is a unique state in which the a tissue is marginally soft but isotropic. 

When $\ptwo > 2 $ the stress' linear response changes. Specifically, the traceless part depends on the trace $\Tr{Q_1^*}={Q_1^*}^\alpha_\alpha $, corresponding to the change of area due to the forcing. The orientation depends on  that of the network tensor $\tilde{Q}^{\mu\nu}$. As such, the actual stress may have a shear component (off-diagonal), even though the strain has only normal components. Additionally, the normal response's sign depends on the orientation of $\tilde{Q}^{\mu\nu}$. 

A calculation of  the complex $\Tr{Q_1^*}^{\mu\nu}$ gives:
\begin{align} \label{eq: Q1trace}
	\Tr{Q_1^*} = &  \frac{- n_1 \omega^2 - i n_2+ i n_3 \omega}{ d_1 - d_2 \omega^2 + i d_3 \omega} \\  \nonumber
	d_1 =& 4\left(1+2H\right) k_P \left[\tilde{Q}^{11}-\tilde{Q}^{22} + 4 \left(\tilde{Q}^{12}\right)^2\right]\\ \nonumber
	d_2 =& 4 \\ \nonumber
	d_3 =& 4 \left[1+2H \left(1+ k_P \ptwo \Tr\tilde{Q}\right)+2k_P\left(\Tr^2 \tilde{Q}-2\right)\right] \\ \nonumber
	n_1 =& 2 k_P\left(\tilde{Q}^{11}-\nu \tilde{Q}^{22}\right) \\ \nonumber
	n_2 = & k_P\left(1+2H\right) \left(1+\nu\right)  \tilde{Q}^{12}\Tr \tilde{Q} \\ \nonumber
	n_3 = &k_P\left(1+2H\right) \left[\left(1+\nu\right)\tilde{Q}^{11}\tilde{Q}^{22} \Tr \tilde{Q}-2 \left(\tilde{Q}^{12}\right)^2 \left(\tilde{Q}^{11}-\nu \tilde{Q}^{22}\right)\right].
\end{align}
This expression depends on the $\tilde{Q}^{\mu\nu}(\nu,\ptwo)$.

\begin{figure*}[h!]
	\centering
	\includegraphics[width=\textwidth]{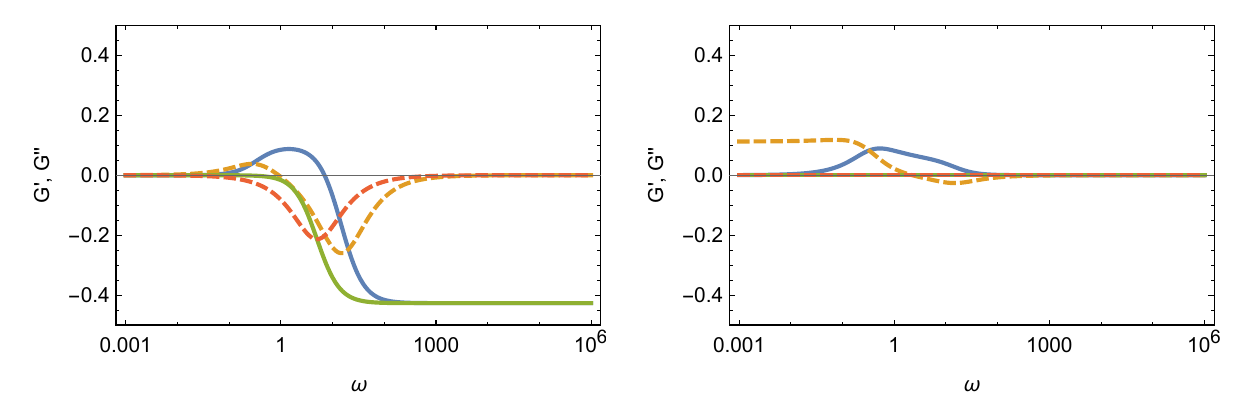}
	\caption{\label{fig: linear rheology} Real and imaginary parts of the complex modulus $G'_n$ (solid lines) and $G''_n$ (dashed lines) for a tissue on a substrate with $\ptwo=3>2$. Left - when $\nu=0<\nu^*$. Blue and Yellow - when $H=0$, Green and Orange  - when $H=-\frac{1}{2}$. Notice the crossover when $H>-\frac{1}{2}$, as a result of the change in cell area relative to the substrate. Right, same as before, when $\nu = \frac{1}{2}> \nu^*$. In the case $H=-\frac{1}{2}$ the modulus vanishes, indicating no change of area relative to the substrate (and only a rotation).  In all plots $k_P=1$ }
\end{figure*}
\begin{figure*}[h!]
	\centering
	\includegraphics[width=.5\textwidth]{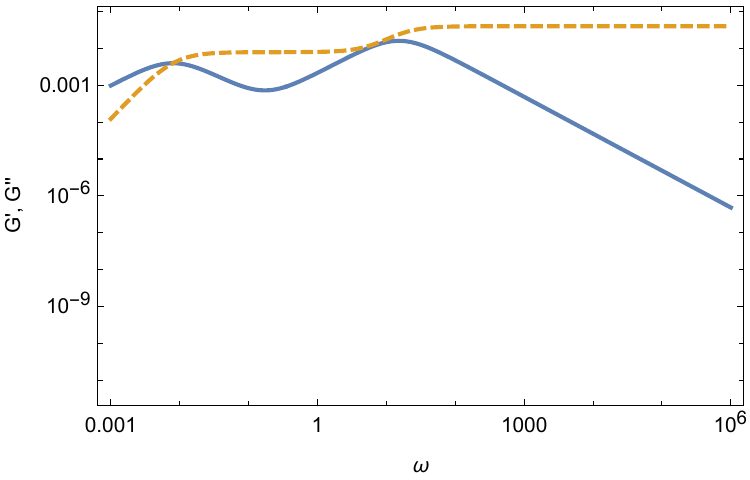}
	\caption{ Real and imaginary parts of the complex modulus $G'_s$ (solid lines) and $G''_s$ (dashed lines) for a tissue under shear  with $\ptwo=2.1>2$. Results similar to numerics in \cite{tong2023linear} }
\end{figure*}
\clearpage	
\section{Appendix G - Fast Oscillations }
Here we plot the area, perimeter, angle energy, and the the $\Qtwo^{12}$ bot actual and on the reference state, of a supported tissue on a fast oscillating substrate, as a function of the substrate's Poisson's ratio.

\begin{figure*}[!h]
	\centering
	\includegraphics{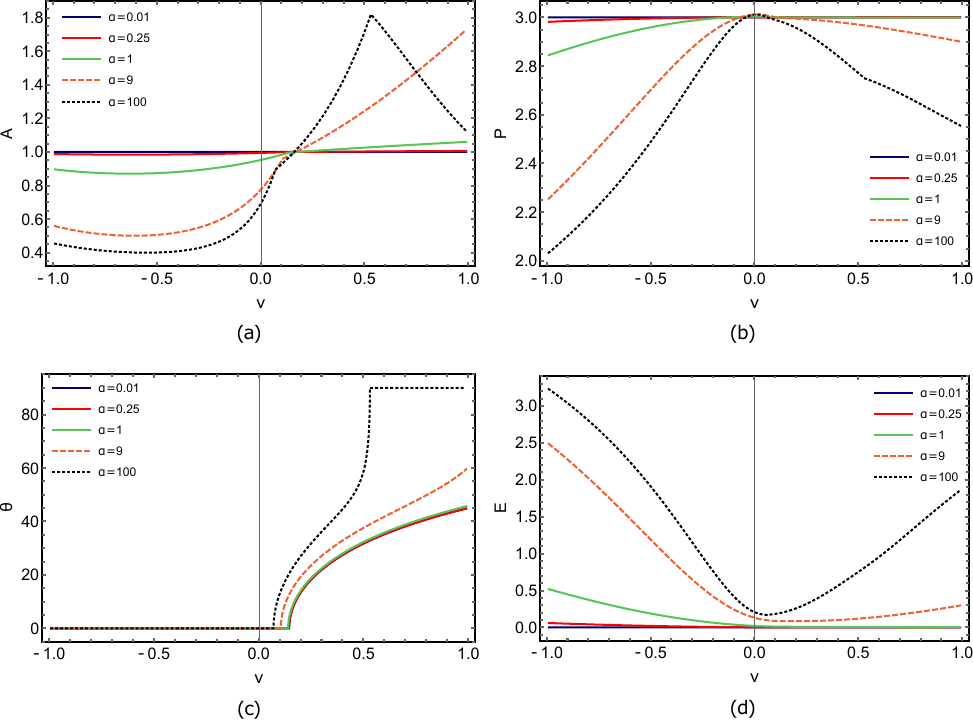}
	\caption{Average tissue area (a), perimeter (b), angle (c), and energy (d), for compatible tissues supported on an oscillating elastic medium, as a function of the material's Poisson's ratio, for different oscillation amplitudes. \label{fig: oscillation all}}
\end{figure*}
\begin{figure*}[!h]
	\centering
	
	\includegraphics{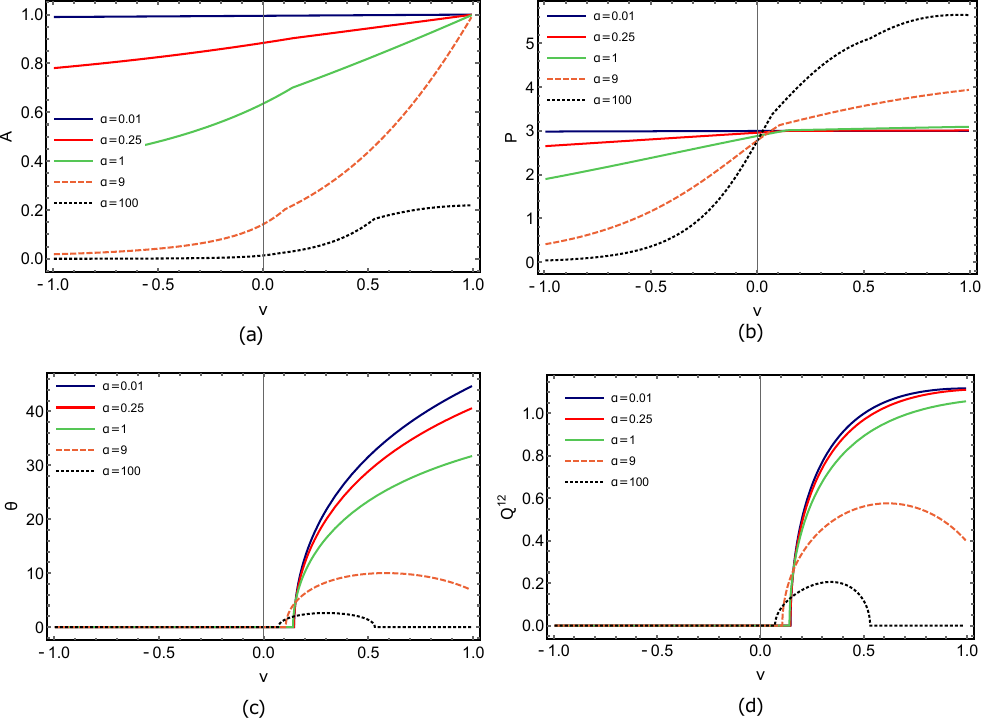}
	\caption{Actual tissue area (a), Perimeter (b), angle (c), and $\Qtwo^{12}$ (d), for compatible tissues supported on an oscillating elastic medium, at the moment when oscillations stop, as function of the material's Poisson's ratio, for different oscillation amplitudes. \label{fig: oscillation all_stop}}
\end{figure*}


\clearpage
\section{Appendix H - Tricritical point }
Here we show the cells angle as a function of the substrate Poisson's ratio, at different frequencies. 
\begin{figure}[h!]
	\includegraphics{{Above and after_j.pdf}}
	\includegraphics[width=0.7 \columnwidth]{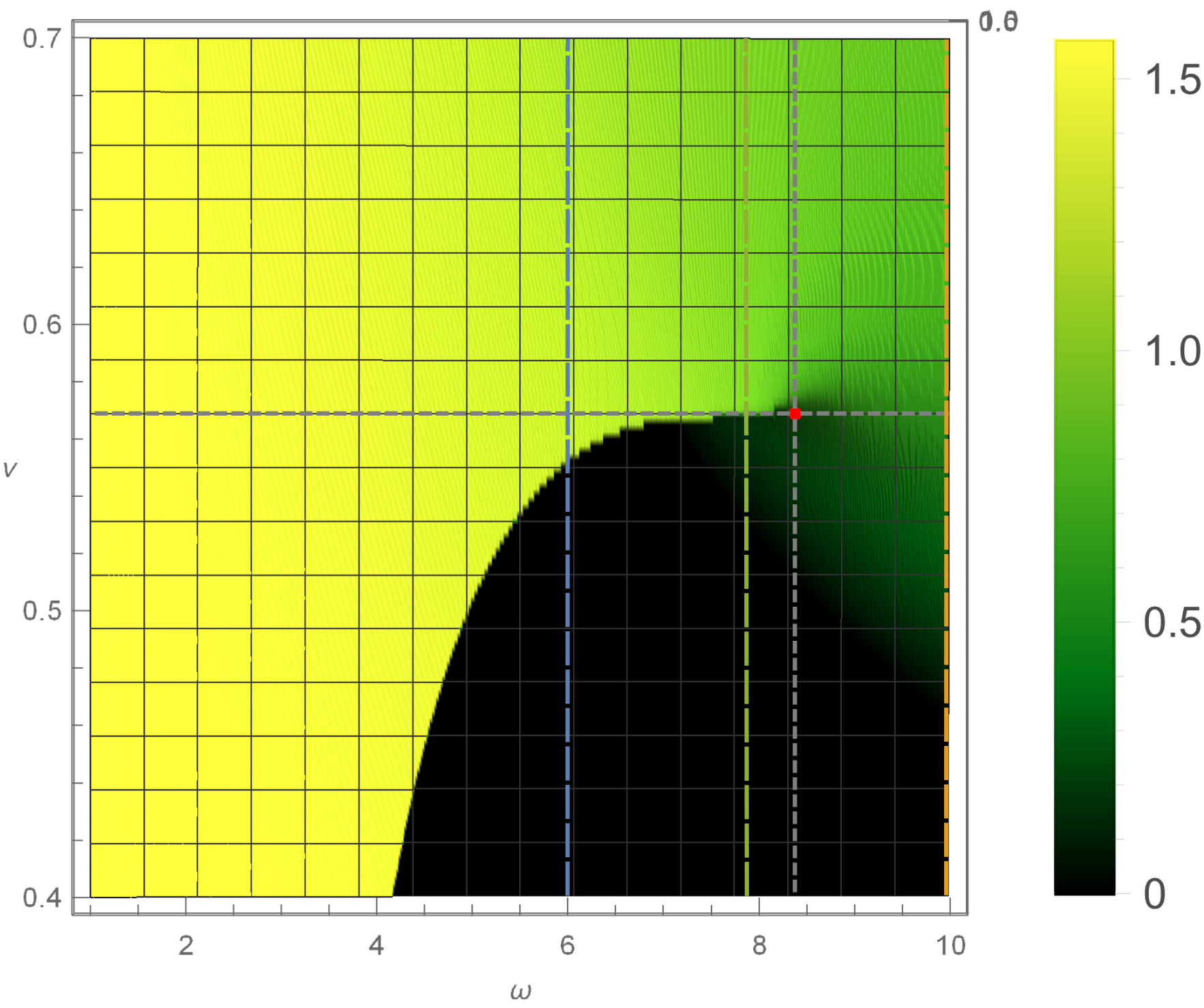}
	\caption{Upper  -Equilbrium angle $\theta$ as a function of substrate's Poisson's ratio $\nu$ at $\omega =6 < \omega_c$ (dashed blue), $\omega = 7.8< \omega_c$ (dashed-dotted green), and $\omega=10>\omega_c$ (yellow). Light, dotted lines, indicate a sudden jump.  Below $\omega_c$ the transition is abrupt (at $\nu \sim 0.55$), indicated a - first order transition. Above $\omega_c$ the transition is continuous (at $\nu \sim 0.46$), but not smooth. Near the transition, a continuous transition begins (at $\nu \sim 0.53$), followed by an abrupt jump at ($\nu \sim 0.57$); Lower -  same as Fig.7 in main text, colored lines - position of the curves calculated above (same colors) \label{fig: tricrit} }
\end{figure}
	
\clearpage
\bibliography{D:/Users/grossman/Documents/PostDoc_Old/Projects/Papers_and_Notes/bibs/active}